\documentclass[aps,cyrwin,prb,twocolumn,floats,graphicx,epsfig,euscript,amsmath,amssymb]{revtex4}
\usepackage{mathrsfs}
\usepackage[dvips]{graphicx}
\usepackage{epsfig}
\DeclareMathAlphabet\EuScript{U}{eus}{m}{n} \SetMathAlphabet\EuScript{bold}{U}{eus}{b}{n}

\def\lapprox{\,\raise0.4ex\hbox{$<$}\kern-0.8em\lower0.7ex\hbox{$\sim$}\,}
\def\gapprox{\,\raise0.4ex\hbox{$>$}\kern-0.8em\lower0.7ex\hbox{$\sim$}\,}
\def\ba{\begin{array}}
\def\ea{\end{array}}

\begin{document}
\title
{Damping via the hyperfine interaction of a spin-rotation mode in a two-dimensional strongly magnetized electron plasma }

\author{$\qquad$ S. Dickmann}

\affiliation{$$Institute of Solid State Physics, Russian
Academy of Sciences, Chernogolovka, 142432, Russia}

\date{\today}

\begin{abstract}
 We address damping of a Goldstone spin-rotation mode emerging in a quantum Hall ferromagnet due to laser pulse excitation.  Recent experimental data show that the attenuation mechanism, dephasing of the observed Kerr precession, is apparently related not only to spatial fluctuations of the electron Land\'e factor in the quantum well, but to a hyperfine interaction with nuclei, because local magnetization of GaAs nuclei should  also experience spatial fluctuations. The motion of the macroscopic spin-rotation state is studied microscopically by solving a non-stationary Schr\"odinger equation. Comparison with the previously studied channel of transverse spin relaxation (attenuation of Kerr oscilations) shows that relaxation via nuclei involves a longer quadratic stage of time-dependance of the transverse spin, and, accordingly, an elongated transition to a linear stage, so that a linear time-dependance may not be revealed.
\end{abstract}
\maketitle
\section{Introduction}

Observation of Kerr rotation in a magnetized two-dimensional electron system (2DES) reveals a long-living spin-rotation mode emerging due to specific laser pulse action.\cite{Fukuoka,la15} This mode macroscopically corresponds to precession of the total spin if it is tilted by an angle with respect to the magnetic field direction. At filling factor $\nu\!=\!1$ the precession rotation with frequency $\epsilon_{\rm Z}/h$ ($\epsilon_{\rm Z}\!=\!g\mu_BB$ is the Zeeman energy) attenuates in characteristic time $T_2\!\gtrsim\! 10\,$ns; however the damping time falls if $\nu$ deviates from unit (e.g., if $|\nu\!-\!1|\!\gtrsim\!0.1$), becoming of the order $\!1\,$ns. Kerr rotation monitoring is a convenient tool and so far probably the only way to investigate spin stiffness of such quantum Hall (QH) magnetics at different filling factors and temperatures.\cite{ku20}
The discussion presented in the recent work $\,$\cite{di20} shows that, in general, a macroscopic approach can be applied to a QH ferromagnet (similar to the study of the Kerr rotation  effect in common dielectric ferromagnets),  however, in this case the Landau-Lifshitz equation hardly is applicable. This is due to the existence of an elementary stochastization process in the magnetized 2DES which saves both Zeemann and exchange energies but results in decreasing total spin $S$,  component $S_z$ remaining unchanged. The stochastization mechanism is determined by smooth spatial fluctuations of the $g$-factor in the 2DES, and has a simple physical meaning: within the single-electron approach, the electrons do not precess coherently but with slightly different Larmor frequencies in different places of the 2D space. { Such } spatial $g$-factor disorder could be related, for instance, to  the stochastic nature of the doping distribution,\cite{kuh17} or/and spatial fluctuations of the quantum well thickness.\cite{iv07}

The present research is devoted to another mechanism that also contributes to spin-rotation stochastization, namely, the influence of correlated spatial fluctuations of nuclear spins interacting with 2DES electrons. This influence can be considered in terms of an addition to the effective electron Zeeman energy. Recent experimental data $\,$\cite{ku20} show that under some conditions this mechanism of hyperfine coupling with correlated nuclei is the main one, or at least can compete with that considered before.\cite{di20}

The reason for the appearance of the spatial correlation of nuclear spins in the quantum well is related to the features of the experimental technique. Indeed, the Goldstone spin-rotation mode is the result of { long laser treatment} of the 2DES (up to 20-30 minutes) providing continuous pumping of electron--heavy-hole pairs into the system. The vast majority of these `laser-induced' electrons immediately recombine with their `gemini' valence holes (see footnote 4 in Ref.\onlinecite{di20}), however, owing to long pumping time, they still disrupt the distribution of nuclear spins via a contact interaction with the nuclei. The latter were spatially uncorrelated in the thermodynamic equilibrium state before laser treatment. (It is more correct to say that the correlation length for nuclear spins in the equilibrium state is of the order of the lattice constant.)
In turn, the dynamic spatial laser-induced distribution of electrons with spins directed along or against the laser beam direction$\,$\cite{ku20} is determined by the fluctuations of the electromagnetic field polarization/amplitude in the cross section of the beam. The relevant spatial characteristic of these fluctuations is of the order of the light wavelength. Thus, we assume that, as a result, the distribution of nuclear spins in the 2D space acquires long-range ($\Lambda\!\sim 300\,$nm) correlations, although the nuclear spin system as a whole remains almost unpolarized.

In conclusion of this section, the old mechanism$\,$\cite{di20} and the new one studied in this paper have in common that in both situations the cause of spin stochastisation in the 2DES is an external static magnetic disorder. Indeed, in the first case the characteristic time of magnetic disorder is the time of existence of the heterostructure (in fact, years). In the second case, it is the time of nuclear spin relaxation depending on experimental conditions but in the actual study$\,$\cite{ku20} it obviously constitutes at least tens of minutes. Both times are significantly longer than the electron spin relaxation times of the system so that the impact on the electron spins in the present work can effectively again be reduced to frozen spatial fluctuations of the single-electron Zeeman energy. This is the reason why many intermediate results obtained in the general approach$\,$\cite{di20} are also used in the current work. These results that are necessary for developing the theory of the spin-rotation-mode damping due to the hyperfine interaction with nuclei are summarized in the next section. The third section is devoted directly to the theory of  the damping caused by the hyperfine interaction mechanism. The calculation results of the characteristic spin-rotation damping times are given in the fourth section. The fifth section discusses these results in the light of newly obtained experimental data.

\section{Description of spin-rotation mode. Formalism of  `excitonic representation'.}

 The QH spin-rotation state is macroscopically equivalent to deviation by the angle of the direction of total spin ${\bf S}$ with conserved value ${\bf S}^2$, and is quantum-mechanically presented as $\,$\cite{la15,di20}\vspace{-1.mm}
\begin{equation}\label{GM}
|\{C_n\},t\rangle=e^{-iE_0t}\sum_{n=0}^{{\cal N}_\phi}C_n e^{-in\epsilon_{\rm Z} t}(S_-)^n|0\rangle.\vspace{-1mm}
\end{equation}
 Here $|0\rangle$ is considered to be the QH odd-integer ground state, where all the electrons on the highest (but not empty) Landau level have spins aligned along magnetic field ${\bf B}\!\parallel\!{\hat z}$; ${\cal N}_\phi$ is the Landau level degeneration number; $S_-$ {describes } operator $S_x\!-\!iS_y$. Note that, generally, the magnetic field is tilted by an arbitrary angle relatively to the normal to the 2D heterostructure, so the $({\hat x},{\hat y},{\hat z})$ spin space differs from the coordinate space $({\hat X},{\hat Y},{\hat Z})$ used to describe the spatial states of the 2D electrons. State \eqref{GM} represents a solution $|t\rangle$ of the non-stationary Schr\"odinger equation:
 \begin{equation}\label{nonSE}
 i\partial|t\rangle/\partial
t\!=\!{\hat H}_0|t\rangle
\end{equation}
(assuming $\hbar\!=\!1$) with the Hamiltonian commuting with the $S_z$ and ${\bf S}^2$ operators, precisely\vspace{-2mm}
\begin{equation}\label{Ham0}
{\hat H}_0\!=\!-\epsilon_{\rm Z}S_z+H_{\rm int}, \vspace{-1mm}
\end{equation}
where $H_{\rm int}$ describes all the spinless
interactions including the Coulomb correlations and single-particle interactions determined by the spin-independent external fields. We remind that in this paper, as in the previous theoretical study$\,$\cite{di20} and in the experiments$\,$\cite{Fukuoka,la15,ku20}, a QH ferromagnet is studied, i.e. a strongly correlated multi-electronic system in which the Coulomb interaction is `ab initio' taken into account. The Coulomb coupling is included in the Hamiltonian ${\hat H}_{\rm int}$ and, as a result, there are no free electrons in the system -- the relevant excited states are purely electronic collective excitations called Goldstone and spin-wave excitons (see below). Of course, the Coulomb interaction itself cannot determine any spin relaxation mechanism, since it is spinless (its Hamiltonian commutes with any spin operator). At the same time, the Coulomb interaction plays a fundamental role in describing the states of any QH system. In particular, it (or rather its exchange part) determines the spin-wave spectrum (see value ${\cal E}_q$ below) and the so-called global `spin stiffness' of the QH ferromagnet.

Thus, state \eqref{GM} is an expansion over the basic set of stationary states $|n\rangle\!=\!(S_-)^n|0\rangle$  satisfying equation
\begin{equation}\label{StEq}
{\hat
H}_0|n\rangle\!=\!(E_0\!+\!n\epsilon_{\rm Z})|n\rangle,
\end{equation}
where $E_0$ is the ground state energy.

State \eqref{GM} is given by a set of factors $\{C_n\}$ (assumed to satisfy the normalization rule $\sum_n |C_n|^2\langle n|n\rangle\!=\!1$) that in turn are determined by specific conditions resulting in excitation of the spin-rotation mode at initial time $t\!=\!0$.  A specific set modelling the initial state induced by laser excitation was studied in work \onlinecite{di20}. The macroscopic deviation angle from the ${\hat z}$ direction is $\theta\!=\!\pi/2\!-\arctan{(\langle S_z\rangle\!/S)}$ where $S\!=\!{\cal N}_\phi\!/2$ but $\langle S_z\rangle$ is found by calculation of the expectation { value}: $\langle S_z\rangle=\!\langle t,\{C_n\}|{\hat S}_z|\{C_n\},t\rangle\!$\vspace{-1.mm}
$${}\!\!{}\!\!{}\!\!{}\!\!\ba{l}=\!{N_\phi\!/2-\!\!\sum\limits_{n=1}^{{\cal N}_\phi}n\langle n|n\rangle|C_n|^2}\!.
\vspace{-1.mm}
\ea
$$

We could certainly imagine a situation where the scenario of damping of state \eqref{GM} would depend on specific set $\{C_n\}$. However, at least in two important cases the damping represents a process independent of the chosen set $\{C_n\}$. This evidently takes place when (i) decay of any state $|n\rangle$ occurs in the same way independently of the number $n$; or (ii) the $C_n$ numbers have a sharp maximum in the vicinity of $n\!=\!m$, then one can simply study the decay of the only state $|m\rangle$. Both conditions were satisfied in the previous study,\cite{di20} and now we will see that here we have a similar situation, i.e.,
the damping mechanism in question is related to a perturbation that couples state $|n\rangle$ where the spin numbers take exact values $S\!=\!{\cal N}_\phi\!/2$ and $S_z\!=\!{\cal N}_\phi\!/2\!-n$,  with states $|{\bf
q};n\rangle\!$ corresponding to exact spin numbers
$S\!=\!{\cal N}_\phi\!/2\!-\!1$ and $S_z\!=\!{\cal N}_\phi\!/2\!-n$; i.e. the transition $|n\rangle\to|{\bf
q};n\rangle$ is associated with change $\delta S\!=\!-1$ when saving the $S_z$ spin component, where ${\bf q}$ is the 2D nonzero wave-vector of the electron system.

To define the $|{\bf q};n\rangle$ state we, as in the work \onlinecite{di20}, use the `excitonic representation' technique (see Refs. \onlinecite{di20,di19,di12} and the references therein) employing, in particular, ${\cal Q}$-operators:\cite{dz83}\vspace{-1.mm}
 \begin{equation}\label{Q}
{\cal Q}_{\bf q}^{\dag}=\sum_{p}
  e^{-iq_x\!p}\;
  b_{p+\frac{q_y}{2}}^{\dag}a_{p-\frac{q_y}{2}},\vspace{-2.mm}
\end{equation}
where $q$ is dimensionless (in $1/l_B$ units, $l_B$ is the magnetic length), and $a_p/b_p$ are the spin-up\!/-down operators annihilating the electron in the orbital state $p$ of the $l$-th  incompletely occupied Landau level. Then the relevant state is
\begin{equation}\label{q-state}
|{\bf q};n\rangle\!=\!{\cal Q}_{\bf q}^\dag(S_-)^{n\!-\!1}|0\rangle,\vspace{-1mm}
\end{equation}
whereas the unperturbed ground state of the $\nu=2l\!+\!1$ QH ferromagnet is  \vspace{-1.mm}
\begin{equation}\label{2.1}
|{0}\rangle\!=\!a^\dag_{p_1}a^\dag_{p_2}...a^\dag_{p_{{\cal N}\!\!{}_\phi}}|{\rm vac}\rangle,\vspace{-1.mm}
\end{equation}
where $|{\rm vac}\rangle$ represents completely occupied lower Landau levels (numbers $p_j\!=\!2\pi jl_B/L$ labelling the states of the degenerate Landau
{level run } over values $2\pi l_B/L,\, 4\pi l_B/L,...\,2\pi\!{\cal N}_\phi l_B/L\!\equiv\!L/l_B$, where $L\times L$ is dimension of the 2D system).
The ${\cal Q}$-operators \eqref{Q} of a varied kind [where $a_p$ and $b_p$ could belong to different Landau levels or to the same spin states of different or even the same Landau level (see below definition \eqref{intra-spin} of the ${\cal A}$ and ${\cal B}$ operators)] have a very important property: when acting on the state of the QH system, they add value ${\bf q}$ to the total momentum of the system since there occurs a commutator equality \vspace{-1.mm}
\begin{equation}\label{P_commutator}
\left[{\hat P},{\cal Q}_{{\bf q}}^{\dag}\right]={\bf q}{\cal Q}_{{\bf q}}^{\dag},\vspace{-1.mm}
\end{equation}
for ${\hat P}$ describing the dimensionless (with $\hbar=l_B=1$) `momentum' operator.\cite{di19,Gorkov,ka84} In particular, if $|0\rangle$ is the ground state, then the exciton state ${\cal Q}_{{\bf q}}^{\dag}|0\rangle$, if not zero, is the eigenstate of momentum operator ${\hat P}$ with  eigen quantum number ${\bf q}$. Thus, exciton states, in contrast to single electron states, possess a natural quantum number, namely, the 2D momentum whose existence is the consequence of the translational invariance of the system. If we take into account the equivalence
\begin{equation}\label{ba-inverse}
  b^\dag_{p_2}a_{p_1}\equiv\sum_{\bf
  q}\frac{e^{iq_x(p_2\!-\!q_y/2)}}{{{\cal N}_\phi}}\delta_{q_y,p_2\!-\!p_1}{\cal Q}^\dag_{\bf
  q}
\end{equation}
(here $\delta_{...,...}$ is the Kronecker delta),  any interaction can be expressed in terms of ${\cal Q}$-operators of one kind or another depending on the correct choice of operators $a_p$ and $b_p$ corresponding to the interaction type. In particular, the unperturbed Hamiltonian ${\hat H}_0$ including the $e$-$e$ Coulomb coupling is presented in terms of excitonic representation operators.\cite{di20}

The key point of the study is as follows. There are two eigen vectors $|n\rangle$ and $|{\bf
q};n\rangle$ of the ${\hat H}_0$ Hamiltonian, and, in spite of the formal operator equivalence ${\cal Q}_{\bf 0}^\dag\equiv S_-$, these represent different spin
states. Indeed, there is no continuum transition from the
${\bf q}\!\equiv\!0$ and ${\bf q}\!\to\!0$ states.\cite{di20}
In particular, for instance, at {\em any} {\bf q} (even in the infinitesimally small limit, $q\to 0$) the `spin-wave exciton' ${\cal Q}_{\bf
q}^\dag|0\rangle$ changes the total spin numbers equally by $\delta S=\delta S_z=-1$ as compared to the ground state
(this fact can be checked straightforwardly by the action of the $S_z$ and ${\bf S}^2$ operators onto spin-wave state ${\cal Q}_{\bf
q}^\dag|0\rangle$), whereas the `Goldstone exciton' $S_-|0\rangle$ corresponds to change by $\delta S_z=-1$ and $\delta S=0$.
The stochastization process, the spin-rotation dephasing, is a transition from $|n\rangle$ to state $\left|{\bf
q};n\rangle\right|_{{\bf
q}\!\to 0}$ having the same energy. The transition occurs in the presence of perturbation mixing these states. This can be an operator of the form \vspace{-1.mm}
\begin{equation}\label{operator}
{\hat V}={{\cal N}_\phi}\!{}^{-1/2}\!\sum_{{\bf q}\neq 0}\!{\cal F}({\bf q})(\!{\cal A}_{\bf q}^\dag\!-\!{\cal B}_{\bf q}^\dag), \vspace{-1.mm}
\end{equation}
studied in the work \onlinecite{di20}. Here ${\cal A}_{\bf q}^{\dag}$ and ${\cal B}_{\bf q}^{\dag}$ are the intra--spin-sublevel operators: \vspace{-1.mm}
\begin{equation}\label{intra-spin}
{\cal A}_{\bf q}^{\dag}\!=\!\sum_{p}
e^{-iq_x\!p} a_{p+\frac{q_y}{2}}^{\dag}a_{p-\frac{q_y}{2}}\vspace{-2.0mm}
\end{equation}
(${\cal B}_{\bf q}^{\dag}$ means the $a\to b$
substitution) which satisfy the following commutation rules:
$$\begin{array}{l}
e^{i{\bf q}_1\!\times{\bf q}_2/2}\!\!\left[{\cal A}_{\bf q_1}^\dag,
  {\cal Q}_{{\bf q_2}}^\dag\right]\!=\!
  -e^{-i{\bf q}_1\!\times{\bf q}_2/2}\left[{\cal B}_{\bf q_1}^\dag,
  {\cal Q}_{{\bf q_2}}^\dag\right]\!\\\qquad\qquad{}\qquad{}\qquad\;=
  -{\cal Q}_{{\bf q_2\!+\!q_1}}^\dag
\end{array}\vspace{-2mm}
$$
and\vspace{-2mm}
$$
\displaystyle{\left[{\cal Q}_{{\bf q_1}}\!,
  {\cal Q}_{{\bf q_2}}^{+}\right]\!\!=\!
  e^{i{\bf q}_1\!\times{\bf q}_2/2}\!\!{\cal A}_{\bf q_1\!-
  \!q_2}\!\!-\!e^{-i{\bf q}_1\!\times{\bf q}_2/2}{\cal B}_{\bf q_1\!-
  \!q_2}} \vspace{-1mm}
$$
Then the formal solution of the non-stationary Schr\"odinger equation $i\partial|t\rangle\!/\partial
t\!=\!({\hat H}_0+V)|t\rangle$ [cf. Eq. \eqref{nonSE}; see Eqs. \eqref{Ham0} and \eqref{operator}]  by means of the manipulations given in work \onlinecite{di20}, allows to find  the initial evolution of the $S\!{}_\perp$ spin-component perpendicular to the ${\bf B}$ direction $\left[S\!{}_\perp(t)^2={\bf S}(t)^2-S_z^2\right]$:\vspace{0mm}
\begin{equation}\label{S_perp}
S\!{}_\perp(t)\!=S\!{}_\perp(0)e^{-i\epsilon_{\rm Z}t}\left[1\!-\!f(t)\right].\vspace{0mm}
\end{equation}
This formula is valid if $f(t)\ll 1$,  where function $f$ is defined by the expression\vspace{-1.mm}
\begin{equation}\label{f}
f=\displaystyle{4\!{}\!\int_0^{{\cal E}\!\!{}_\infty}}
\!\!|{\cal F}({\bf q})|^2\!\left[1\!-\!\cos{\!({\cal E}_qt)}\right]\!{\nu({\cal E}_q\!)d{\cal E}_q}\!/{{\cal E}_q}\!\!{}^2,\vspace{-1.mm}
\end{equation}
Here ${\cal E}_q$ is the $q$-dispersion of spin-wave excitons in the system first studied theoretically in works \onlinecite{by81} and \onlinecite{ka84}, and experimentally  in works \onlinecite{pi}. For our purposes, only the initial quadratic part of the dispersion curve corresponding to small values of $q$ is significant: namely, if $q\ll 1$ then
\begin{equation}\label{dispersion}
{\cal E}_q\!\approx\!q^2/2M_{\rm x},
\end{equation}
where $M_{\rm x}$ is the spin-wave mass whose magnitude is inversely proportional to the characteristic Coulomb energy $e^2/l_B$,\cite{ka84,by81}  but experimentally it is found that $M_{\rm x}\!\approx\!0.5\,$meV${}^{-1}\,$ in the relevant range of magnetic fields.\cite{pi} $\nu({\cal E}_q\!)$ denotes the density of the spin-wave states,
$\nu(0)\!=\!M_{\rm x}$, in particular. Eqs. \eqref{S_perp} and
\eqref{f} were derived in work \onlinecite{di20} where the perturbation operator \eqref{operator}, i.e., the function ${\cal F}$, was determined by smooth spatial fluctuations of the electron $g$-factor in the 2DES.

Now let us study a different type of perturbation, and we will see that the problem actually  boils down to finding another function ${\cal F}$.

\section{Hyperfine interaction of conduction electrons with nuclei as the reason for smooth spatial fluctuations of effective Zeeman energy}

We consider the hyperfine contact interaction of conduction electrons with nuclei,\cite{abr} starting directly from the well known expression (see, e.g., Ref. \onlinecite{di12} and the references therein): ${\hat V}_{\rm hf}\!=\!\sum_i{\hat V}_{\rm hf}^{(i)}$, where\vspace{-1.mm}
\begin{equation}\label{HC}
 {\hat V}_{\rm hf}^{(i)}=\frac{v_0}{2}A_i\Psi^*(\mbox{\boldmath
 $R$}_i)\left(\hat{\mbox{\boldmath $I$}}^{(i)}\!\!\!\!\cdot\hat{\mbox{\boldmath
 $\sigma$}}\right)\Psi(\mbox{\boldmath $R$}_i)\,.\vspace{-1.mm}
\end{equation}
Here $\hat{\mbox{\boldmath $I$}}^{(i)}$ and $\mbox{\boldmath $R$}_i$ are the spin operator and position of
the $i$-th nucleus, $\Psi(\mbox{\boldmath $R$})$ is the electron envelope function
[$\mbox{\boldmath $R$}=(X,Y,Z)$ is the 3D vector], $\hat{\mbox{\boldmath
 $\sigma$}}$ are the Pauli matrixes. We study the case of a GaAs semiconductor
 structure, hence, both nuclei have the same
total spin: $I^{\rm Ga}\!=\!I^{\rm As}\!=\!3/2$, and as value $v_0$ we consider the volume of the unit cell containing two atoms: $v_0\!=\!45.2\,$\AA${}^3$. The parameters $A_i$ are proportional to the Ga/As nuclear magnetic moments. Both $A_{\rm Ga}$ and $A_{\rm
As}$ depend only on the positions of the Ga/As nuclei within the unit cell. For the final calculation we need the sum  $A_{\rm GaAs}=\overline{A_{\rm Ga}}\!+\!A_{\rm
As}\!\simeq\!1\,$K calculated for the case when the electron Bloch function is normalized within volume $v_0$. (The overline means averaging over the isotopic composition of the gallium atoms in the GaAs lattice; see Appendix A in Ref. \onlinecite{di12}.)

Hamiltonian \eqref{HC} is an operator acting on electrons and nuclei. First, we transform it focusing on its role in the $|n\rangle\!\to\!|{\bf q};n\rangle$ elementary transition.
If $\hat{\mbox{\boldmath $I$}}^{(i)}\!\!\!\!\!\cdot\,\hat{\mbox{\boldmath
$\sigma$}}$ is rewritten as ${\hat I}_z{\hat \sigma}_z\!+\!{\hat I}_+{\hat \sigma}_-\!+\!{\hat I}_-{\hat
\sigma}_+$, then it becomes clear that, unlike the case of relaxation changing the $S_z$ component,\cite{foot1} here we have to leave only Ising term
${\hat I}_z{\hat \sigma}_z$ ensuring stochastization of Kerr precession at conserved $S_z$.
In the 2D channel we have $\Psi(\mbox{\boldmath $R$})\!=\!\chi(Z)\psi(X,Y)$ where $\psi(X,Y)\!=\!\sum_p[a_p\psi_{l, p,\uparrow}(X,Y)\!+b_p\psi_{l, p,\downarrow}(X,Y)]$ ($\psi_{l,p,\sigma}$ are the $l$-th Landau-level wave functions), and substituting the Schr\"odinger operators ${{\hat
\Psi}^\dag}\!$/${{\hat \Psi}}$ for ${\Psi}^*\!$/${\Psi}$ we come to\vspace{-1.mm}
\begin{equation}\label{HF2}
\begin{array}{l}
\!\!\displaystyle{{\hat V}_{\rm
hf}=\frac{v_0}{2}\sum_{p_1,p_2}(a^\dag_{p_2}a_{p_1}\!-b^\dag_{p_2}b_{p_1})}\\
\!\!\!\!\!\displaystyle{\times\!\!\sum_i|
\chi(Z_i)|^2\psi_{l,p_2}^*\!(\!X_i,\!Y_i)\psi_{l,p_1}\!(\!X_i,\!Y_i)A_i{\hat
I}_z^{(i)}}.
\end{array}\vspace{-1.mm}
\end{equation}
Expressing in terms of the exciton operators (e.g., $a^\dag_{p_2}a_{p_1}\!=\!{\cal N}_\phi^{-1}\!\sum_{\bf q}e^{iq_x(p_2-q_y/2)}\delta_{q_y,p_2-p_1}{\cal A}^\dag_{\bf q}$) and then summating over $p_1$ and $p_2$, we obtain\vspace{-1.mm}
\begin{equation}\label{HF3}
\displaystyle{{\hat V}_{\rm hf}\!=\!\frac{1}{2}\sum_{\bf q}\!{e^{-q^2/4}}L_l(q^2\!/2){\hat F}({\bf q})\!\left({\cal A}^\dag_{\bf q}\!-{\cal B}^\dag_{\bf q}\right)}\vspace{-1.mm}
\end{equation}
($L_l$ is the Laguerre polynomial).
This is similar to Eq. \eqref{operator}, however, now
${\hat F}({\bf q})$ is not a function but an operator acting on nuclear spins
\begin{equation}\label{operatorF}
{\hat F}({\bf q})\!=\frac{v_0}{2\pi l_B^2{\cal N}_\phi}\!\sum_i\!A_i|\chi(Z_i)|^2
e^{i{\bf q}\mbox{\boldmath $R$}_i/l_B}\hat{I}_z^{(i)}\!\!.\vspace{-1.mm}
\end{equation}

For further calculation we have to adequately describe the state of nuclei established after termination of laser pumping, that is at moment $t=0$.
It should be taken into account that the nuclear relaxation times are much longer than the times of the studied electron processes.  Generally, the state of
nuclei is $|\phi\rangle=|\phi_1,\phi_2,...,\phi_i,...\rangle$ where $|\phi_i\rangle$ is a four-component spin function (since the number of components is $2I\!+\!1\!=\!4$). However, in fact, the nuclear spin `gets stuck' in some stationary quantum state, i.e. $\phi_i$ is
an eigen state of the $i$-th nucleus,\cite{abr} then ${I}_z^{(i)}|\phi_i\rangle\!=\!m_i|\phi_i\rangle$ and quantum averaging of the vector operator ${\mbox{\boldmath $I$}}^{(i)}$ results only in a nonzero $z$-component: $\langle \phi|{\mbox{\boldmath $I$}}^{(i)}|\phi\rangle\!=\!\langle \phi_i|{\mbox{\boldmath $I$}}^{(i)}|\phi_i\rangle\!=\!(0,0,m_i)$. Now the state of the nuclear system is determined by a set of $m_i$ numbers. So, operator \eqref{HF3} can be rewritten with ${\hat F}$ replaced by function $F({\bf q})\!=\!\langle \phi|{\hat F}({\bf q})|\phi\rangle$, where\vspace{-0.mm}
\begin{equation}\label{F}
{F}({\bf q})\!=\frac{v_0}{2\pi l_B^2{\cal N_\phi}}\!\sum_i\!A_i|\chi(Z_i)|^2
e^{i{\bf q}\mbox{\boldmath $R$}_i/l_B}m_i.\vspace{-1.mm}
\end{equation}

First, in order to summate over all the nuclei in Eq. \eqref{F}, we perform spatial averaging of the $m_i$ numbers inside domain $dV_{\!\mbox{\tiny\boldmath $R$}}\!=\!d{X}dYdZ$ centered at point $\mbox{\boldmath $R$}$. Supposing the domain volume is much larger than $v_0$, but the dimensions are much smaller than the quantum well width, $\sim (\int |\chi(Z)|^4dZ)^{-1}$, in the ${\hat Z}$ direction, and the correlation length $\Lambda$ and wave-length $2\pi l_B/q$ in the $({\hat X},{\hat Y})$ plan, we define the mean value ${\widetilde m}(\mbox{\boldmath $R$})$ of $m_i$ momenta within the domain:
 \begin{equation}\label{m_mean}
 \displaystyle{{\widetilde m}(\mbox{\boldmath $R$})=\frac{v_0}{2}\,(dV_{\!\mbox{\tiny\boldmath $R$}})\!{}^{-1}\!{}\!{}\!{}\!{}\!{}\!{}\!{}\!{}\!{}\!{}\!{}\!{}\!{}
 \sum_{\mbox{\em{i}}\;\; \mbox{within the}\atop \mbox{ domain}\;\;\mbox{${dV}\!$}_{\!\mbox{\tiny\boldmath $R$}}}}\!{}\!{}\!{}\!{}\!{}\!{}\!{}\!{}\!{}m_i\quad.
 \end{equation}
 Second, we consider the value averaged along the quantum well width:
\begin{equation}\label{m_r}
{m}({\bf r})=\int |\chi(Z)|^2{\widetilde m}(\mbox{\boldmath $R$})dZ
\end{equation}
depending only on the coordinates in the plane [${\bf r}$ designates 2D vector ${\bf r}=(X,Y)$], and then present expression \eqref{F} in the form\vspace{-0mm}
\begin{equation}\label{F2}
\ba{r}
\displaystyle{{}\!{}\!{}\!{F}({\bf q})\!\!=\!\!\frac{\overline{A_{\rm Ga}}+A_{\rm As}}{2\pi l_B^2{\cal N}_\phi}\!\!\int\!\!d{\bf r}\,
e^{i{\bf qr}/l_B}{m}({\bf r})}\qquad{}\qquad{}\quad{}\vspace{-2.mm}\\{}\qquad
\displaystyle{\equiv A_{\rm GaAs}{\overline{m}}({\bf q})},
\ea\vspace{-0.mm}
\end{equation}
where
substitution $m({\bf r})=\sum_{\bf q}e^{-i{\bf qr}/l_B}\overline{m}({\bf q})$ has been done. If $m({\bf r})$ has a constant part, i.e. averaging over the space is not vanishing: $\langle m({\bf r})\rangle\!=\!\int m({\bf r})d{\bf r}/L^2\!\neq\!0$, then it does not contribute to the integral in Eq. \eqref{F2} at  ${\bf q}\!\neq\!0$. So, ${\overline{m}}({\bf q})$ is the Fourier component of the spatially fluctuating part $\delta m({\bf r})\!\equiv\!m({\bf r})\!-\!\langle m({\bf r})\rangle$. This $m({\bf r})$-disorder is considered isotropic and, hence, characterized by the correlator:
\begin{equation}\label{correlator}
\int\!\delta m({\bf r}_0)\delta m({\bf r}_0\!+\!{\bf r})d{\bf r}_0\!/L^2\!=\!M(|{\bf r}|),
\end{equation}
where $\sqrt{M(0)}\!\equiv\!\Delta_m$ is the amplitude of the long-distant magnetic disorder, and, besides $M(r)\!\to\! 0$ if $r\!\to\!\infty$. The characteristic range of attenuation of correlator $M(r)$ is equal to correlation length $\Lambda$. (For example, in the case of Gaussian disorder we have $M(r)\!=\!\Delta_m^2e^{-r^2/\Lambda^2}$.) The Fourier component
\begin{equation}\label{М_Fourier}
\overline{M}({ q})\!=\!\int\!M({
r})e\!{}^{-i{{\bf qr}/l_B}}d{\bf r}/(2\pi l_B)^2
\end{equation}
is also a function of the ${\bf q}$ modulus and related to the $\overline{m}({\bf q})$ value with formula $\overline{M}({
q})\!=\!{\cal N}_\phi|\overline{m}({\bf q})|^2/2\pi$. Particularly, in the Gaussian case we get
\begin{equation}\label{corr_gauss_Fourier}
\overline{M}({q})\!=\!\!\displaystyle{{\Delta_m^2\Lambda^2}e^{-(\Lambda
q/2l_B)^2}\!\!\!/{4\pi l_B^2}}.
\end{equation}

Thus, comparing Eqs. \eqref{operator}, \eqref{HF3} and \eqref{F2}, we find that the ${\cal F}$ function squared, included in formula \eqref{f} is equal to
\begin{equation}\label{function_F}
|{\cal F}({\bf q})|^2 = \frac{\pi}{2}A_{\rm GaAs}^2{\overline M}(q)e^{-q^2/2}[L_l(q^2/2)]^2.
\end{equation}
The most puzzling value in this expression is certainly ${\overline M}(q)$. It does not matter that the magnetic disorder is unlikely to be Gaussian (although this assumption can be used for specific estimates).  More significantly, while we have some idea of the magnitude of correlation length $\Lambda$ (see above), we do not have any theoretical ideas about disorder amplitude $\Delta_m$. At the same time, an estimation of this value based on the available experimental data $\,$\cite{ku20} is possible, which enables us to consider $\Delta_m$ constituting several percent of the nuclear spin value, i.e. $\Delta_m\sim 0.01$.

\section{Results}

Even more obvious than in the case of the relaxation mechanism studied in Ref. \onlinecite{di20} is the fact that only values $q\ll 1$ are important in ${\overline M}(q)$ [see. Eq. \eqref{corr_gauss_Fourier}] and thereby in expression \eqref{function_F} and integral \eqref{f}. This feature is the consequence of very large ratio $\Lambda/l_B\sim 30$.
Then, a simple analysis of Eq. \eqref{f}, made by analogy with the one in Ref. \onlinecite{di20}, shows that: (i) first, at small $t\,$ (if $t\!\ll\!\tau_0\!=\!M_x\Lambda^2/l_B^2\!\simeq 0.2-0.8\,$ns) $f(t)$ is quadratic: \vspace{-1mm}
\begin{equation}\label{quadratic}
f(t)\approx(t/\tau_1)^2, \vspace{-3mm}
\end{equation}
where
\begin{equation}\label{tau1}
\tau_1\!=(\Delta_mA_{\rm GaAs})^{-1}\sim 1\,{\rm ns};\vspace{-1mm}
\end{equation}
and (ii) second, at longer times this function becomes  linear, $f\!\approx\! t/T_2$, in the range:
$\tau_0\ll t\ll T_2$, where time $T_2$ depends on the ${\overline M}(0)$ value regardless of the type of disorder:
\begin{equation}\label{T2}
  T_2^{-1}=\pi^2A_{\rm GaAs}^2{\overline M}(0)M_{\rm x}.
\end{equation}

Considering the quadratic stage of $f(t)$ evolution \eqref{quadratic} we notice that characteristic time $\tau_1$ \eqref{tau1} is independent of spin-exciton mass $M_{\rm x}$ and thereby of the Coulomb interaction proportional to $M_{\rm x}^{-1}$. This result is not surprising: indeed, the initial stage is determined by the mismatch of the Kerr rotation in the system consisting only of Goldstone excitons which are not known to interact with each other. However, the time $\tau_1$, as well as transient time $\tau_0$, is significantly longer compared to the corresponding values in Ref. \onlinecite{di20}, which is due to the fact that the nuclear long-distant magnetic disorder turns out to be effectively smoother than the $g$-factor disorder: the correlation length $\Lambda$ (determining transient time $t_0\!\sim\!1/{\cal E}_q$ with $q\sim l_B/\Lambda$) is longer, and the amplitude $\Delta_m$ [determining $\tau_1$, see Eq. \eqref{tau1}] is smaller.

The linear stage in the $f(t)$ dependence is realizable only if $\tau_0\ll T_2$. Using Eqs. \eqref{corr_gauss_Fourier} and \eqref{T2} and, thus, estimating ratio\vspace{-1.mm}
\begin{equation}\label{ratio_tautoT}
\tau_0\!/T_2\sim A_{\rm GaAs}^2\Delta_m^2M_{\rm x}^2(\Lambda/l_B)^4,
\end{equation}
we get that it is not much less than {unity }, or even of the order of unity. Therefore, {\em the specific feature of the transverse spin relaxation mechanism in question is that the linear stage may be actually absent.} This circumstance, in particular, means that it would be impossible to use an approach considering the problem not quantum-mechanically but kinetically  within the {\large $\tau$}-approximation where the logarithmic rate of the relaxation $d\ln{S_\perp}\!/dt$ is independent of time, cf. Ref. \onlinecite{di20}.

If the Gaussian case is considered \eqref{corr_gauss_Fourier},
integration in Eq. \eqref{f} is
performed analytically
with\vspace{-1mm}
\begin{equation}\label{Fq}
{}\!|{\cal F}({\bf q})|^2\!\!\approx
\!\frac{1}{8}\!
\left(\!A_{\rm GaAs}\Delta_m\Lambda/l_B\right)^2
\!\exp{\!(\!M_{\rm x}{\cal E}_q\Lambda^2\!/2l_B^2)}
\end{equation}
and with $\nu({\cal E}_q)\!\approx\!M_{\rm x}$, and ${\cal E}_{\infty}=\infty$. As a result, we obtain an expression for the $f(t)$ function valid for {\em any relationship between time values} $\tau_0$ {\em and} $T_2$,\vspace{-1.mm}
\begin{equation}\label{f_G}
f(t)=\frac{2t}{\pi T_2^{(G)}}\beta(t/\tau_0^{(G)})\,.\vspace{-2.mm}
\end{equation}
Here $\beta(x)=\arctan{(x)}\!-\!(2x)^{-1}\!\ln{(1\!+\!x^2)}$. $\tau_0^{(G)}\!=\!M_{\rm x}\Lambda^2\!/2l_B^2$ and\vspace{-1.mm}
\begin{equation}\label{T2_Gauss}
1/T_2^{(G)}\!\!=\!\pi M_{\rm x}\left(A_{\rm GaAs}\Delta_m\Lambda/2l_B\right)^2\vspace{-1.mm}
\end{equation}
are the corresponding Gaussian disorder expressions for $\tau_0$ and $T_2\!{}^{-1}$. The only condition that the function \eqref{f_G} must obey is its smallness, which always takes place when $t\!\ll\!T_2$. Substituting into Eq. \eqref{T2_Gauss} the values of the parameters whose estimates are given above, we find $T_2\,\simeq\!10\,$ns, which is in agreement with the experimental data.\cite{la15,ku20}

\section{Discussion}

\vskip -2.mm

We discuss the reported results in the context of the recent research data,\cite{ku20} where the measurements of the Goldstone mode stochastization were taken in a wide range of temperatures and filling factors. Regarding fractional fillings, when $\nu$ deviates from unit, it is known that stochastization is sharply accelerated, supposedly
owing to appearance of additional channels related to some soft modes forbidden in the integer QH ferromagnet. We considered only the case of odd integer $\nu$, however, phenomenologically, such softening of the QH ferromagnet (weakening of spin stiffness) can be associated with softening of the spin-wave mode, that is, with an increase in mass $M_{\rm x}$. Indeed, an increase by an order of magnitude (i.e. when $M_{\rm x}$ is $\simeq\!2\,$K) reduces transverse relaxation time [Eqs. \eqref{T2} and \eqref{T2_Gauss}] by the same amount.  Accordingly, it increases transition time $\tau_0$ and definitely makes the linear law of $f(t)$ impossible. In this case the estimate of $T_2^{-1}\!\simeq\!1-2\,$GHz is in good agreement with our study and the experimental data obtained at $\nu\!=\!0.7$ and $T\!=\!4.2\,$K.

What should happen if the temperature rises, for instance, to 10 K? Theoretical estimates and experimental observations show that in this case the electron spin polarization drops significantly even at filling factor $\nu\!=\!1$. Then, apparently, the dependence of polarization on filling factor becomes fairly weak and, as a result, the stochastization time should not significantly depend on $\nu$ either, constituting about $5\,$ns in the wide range $0.7\!<\!\nu\!<\!1.5$.\cite{ku20} At the same time, it was observed that for a fixed fractional filling factor (for example, for $\nu=0.7$) the increase in temperature from 4.2 to $10\,$K significantly slows the decay of Kerr oscillations (by 5-10 times$\,$\cite{ku20}). This effect cannot be explained if the attenuation mechanism is related to $g$-factor spatial fluctuations which are constant over time and independent of temperature. However, within the framework of the studied nuclear disorder stochastization mechanism, such damping weakening  becomes clear. Indeed, local nuclear magnetization and its long-distant spatial fluctuations caused by laser processing of the system preceding Kerr-precession monitoring, certainly depend on the relaxation time of the nuclei. The latter should significantly depend on temperature and at $T\!=\!10\,$K can be characterized by times shorter than 20-30 min., the experimental time of laser processing. Therefore, as the temperature increases, the amplitude of fluctuations of local nuclear moment $\Delta_m$ decreases; and characteristic times $T_2$, and $\tau_1$ become longer.

\vskip -2.mm

\section{Conclusion}

\vskip -2.mm

The presented work is a development of the theory of the spin-rotation Goldstone mode in a quantum Hall ferromagnet.\cite{di20} An additional mechanism of transverse spin relaxation (stochasticization of Kerr oscillations), different from the one considered previously, is proposed. The new relaxation channel is related to spatial fluctuations of local nuclear magnetization in the GaAs matrix. It does not cancel the previously studied mechanism, but is likely to be dominant under specific experimental conditions.\cite{ku20}

The author is grateful to L.V. Kulik for useful discussions, the Russian Science Foundation (Grant No. \#18-12-00246) for support in interpreting new experimental results, and the Russian Foundation for Basic Research for support in performing some theoretical calculations (Grant No. 18-02-01064).


\vspace{-6.mm}


\begin{thebibliography}{99}


\vspace{-7.mm}


\bibitem{Fukuoka} D. Fukuoka, T. Yamazaki, N. Tanaka, K. Oto, K. Muro, Y. Hirayama, N. Kumada, and H. Yamaguchi, Phys. Rev. B {\bf 78}, 041304(R) (2008);
D. Fukuoka, K. Oto, K. Muro, Y. Hirayama, and N. Kumada, Phys. Rev. Lett. {\bf 105}, 126802 (2010).


\bibitem{la15}
A.V. Larionov, L.V. Kulik, S. Dickmann, and I.V. Kukushkin,
Phys. Rev. B {\bf 92}, 165417 (2015).


\bibitem{ku20} A.V. Larionov, E. Stepanets-Khussein, L.V. Kulik, V. Umansky, and I.V. Kukushkin, Sci. Reports {\bf 10}, 2270 (2020).


\bibitem{di20} S. Dickmann, J. Phys.: Condens. Matter {\bf 32}, 015603 (2020).


\bibitem{kuh17} H. Kuhn, J.G. Lonnemann, F. Berski, J. Hu\"ubner, and M. Oestreich, Phys. Status Solidi B {\bf 254}, 1600574 (2017).


\bibitem{iv07}E.L. Ivchenko. {\em Optical Spectroscopy
of Semiconductor Nanostructures} (Springer, 2007).

\bibitem{dz83}
First the ${\cal Q}$-operators were used as applied to the 2D-electron two-component system in works: {A.B. Dzyubenko and Yu.E. Lozovik}, Sov. Phys. Solid State {\bf 25}, 874 (1983); {\it ibid} {\bf 26}, 938 (1984).

\bibitem{di19} S. Dickmann, L.V. Kulik, V.A. Kuznetsov,  Phys. Rev. B {\bf 100}, 155304 (2019).








\bibitem{di12}
{S. Dickmann and T. Ziman}, Phys. Rev. B {\bf 85}, 045318 (2012).


\bibitem{Gorkov}
L.P. Gor'kov and I.E. Dzyaloshinskii, JETP {\bf 26}, 449 (1968);
I.V. Lerner and Yu.E. Lozovik, JETP {\bf 51}, 588 (1980).


\bibitem{ka84}
{C. Kallin and B.I. Halperin}, Phys. Rev. B {\bf 30}, 5655 (1984).


\bibitem{by81}
Y.A. Bychkov, S.V. Iordanskii, and G.M. Eliashberg, JETP Lett. {\bf 33}, 143 (1981);




\bibitem{pi}
Y. Gallais, J. Yan, A. Pinczuk, L.N. Pfeiffer, and K.W. West, Phys. Rev. Lett. {\bf 100}, 086806 (2008);
I.V. Kukushkin, J.H. Smet, V.W. Scarola, V. Umansky, and K. von
Klitzing , Science {\bf 324}, 1044 (2009) [see the Supporting
Online Material: www.sciencemag.org/cgi/content/full/1171472/DC1].


\bibitem{abr}
A. Abragam, The Principles of Nuclear Magnetism. Clarendon Press. ISBN 9780198520146 (1961).


\bibitem{foot1} In principle, there is a relaxation channel of the Goldstone mode determined by the term proportional to ${\hat I}_-{\hat \sigma}_+$. However, this channel is much slower, characterized  by times of the order of microseconds [see S. Dickmann, JETP Lett. {\bf 93}, 86 (2011)].




\end{thebibliography}
\end{document}